\documentclass[global,onecolumn]{svjour}
\usepackage{graphics}
\usepackage{graphicx}
\usepackage{amssymb}
\usepackage{rotating}
\journalname{Astronomy and Astrophysics Review}
\begin{document}

\title{Galactic discrete sources of high energy neutrinos}

\author{W. Bednarek\inst{1}\thanks{\emph{E-mail address:} 
bednar@fizwe4.fic.uni.lodz.pl}, 
G. F. Burgio\inst{2}\thanks{\emph{E-mail address:} 
fiorella.burgio@ct.infn.it}, 
and T. Montaruli\inst{3}\thanks{\emph{E-mail address:} 
teresa.montaruli@ba.infn.it}}

\offprints{G. F. Burgio}     

\institute{Department of Experimental Physics, University of Lodz, 
ul. Pomorska 149/153, 90-236 Lodz, Poland \and
INFN Sezione di Catania, Via S. Sofia 64, I-95123 Catania, Italy \and
Dipartimento di Fisica and INFN Sezione di Bari,
Via Amendola 173, I-70126 Bari, Italy}
\date{Received: date / Revised version: date}

\maketitle

\begin{abstract}
We review recently developed models of galactic discrete sources of 
high energy neutrinos. 
Some of them are based on a simple rescaling of the TeV $\gamma$-ray fluxes 
from recently detected galactic sources, such as, shell-type 
supernova remnants or pulsar wind nebulae. Others present detailed and
originally performed modeling of processes occurring close to compact
objects, i.e. neutron stars and low mass black holes, which are supposed to 
accelerate hadrons close to dense matter and radiation fields.
Most of the models considered in this review optimistically assume that
the energy content in relativistic hadrons is equal to a significant 
part of the maximum observable
power output in specific sources, i.e. typically $\sim 10\%$. This may 
give a large overestimation of the neutrino fluxes. This is the case of 
models which postulate neutrino production in hadron-photon collisions 
already at the acceleration place, due to the likely 
$e^\pm$ pair plasma domination. 
Models postulating neutrino production in hadron-hadron collisions  
avoid such problems and therefore seem to be more promising.        
The neutrino telescopes currently taking data have not detected 
any excess from discrete sources yet, although some models could already 
be constrained by the limits they are providing.  
\end{abstract}

\tableofcontents
\section{Introduction}
\label{sec:intro}

Galactic sources can potentially produce interesting event rates in
neutrino telescopes.
Since they are at shorter distances to the Earth
($\sim 1 \div 10$ kpc) compared to extra-galactic sources, 
the source luminosity required for a galactic source
to produce the same event rate as an extra-galactic one, is orders 
of magnitude smaller (see Sec.~\ref{sec1}). 
Given the large photon luminosities observed from some of the
sources, it is possible to single out interesting candidate
neutrino emitters in the Galaxy.
A rough estimate of the source luminosity required to produce a
certain event rate in a neutrino telescope (Halzen~2003) is shown in 
Sec.~\ref{sec1}.

In order to produce high energy neutrino fluxes, galactic sources must
accelerate particles at sufficiently high energies. Hillas derived
the maximum energy E at which a particle of charge Z can be accelerated,
from the simple argument that the Larmor radius of the particle 
should be smaller than the size of the acceleration region
(Hillas 1984). If energy losses inside sources are neglected, 
this maximum energy E (in units of $10^{18}$ eV) is related to the strength of 
the magnetic field B (in units of $\mu$Gauss) and the size of the
accelerating region R (in units of kpc) by the following relationship:
\begin{equation}
\rm E_{18} \sim \beta~Z~B_{\mu G}~R_{kpc} 
\label{hill}
\end{equation}
\noindent
where $\beta$ is the velocity of the shock wave or 
the acceleration mechanism efficiency.
Hence, the maximum energy up to which particles
can be accelerated depends on the $B R$ product.
Particle acceleration may occur in many candidate sites, with sizes ranging 
from kilometers to megaparsecs.

Most of the models discussed below assume that neutrinos are produced by 
charged pion decays which in turn are produced in hadron-hadron ($pp$)
and hadron-photon (p$\gamma$) collisions. 
As a matter of fact, in order to produce neutrinos with energies of
the order of 10 TeV, the primary hadron energy has to be about two orders
of magnitude larger in the case of $pp$ collisions, and about one order 
of magnitude larger in the case of p$\gamma$ collisions.
Neutrinos produced as decay products of neutrons (which originate in hadronic 
collisions) are usually much less important since they are produced with 
much lower multiplicities (with respect to charged pions from $pp$ collisions)
and typically take only $\sim 10^{-3}$ of the initial neutron energy. 
Therefore, the fluxes of neutrinos from neutron decay are significantly
lower than those produced in hadronic collisions (through meson decays) 
due to the steepness of the spectrum of accelerated hadrons 
(described usually by a $E^{-2}$ power law).
Only in some special conditions interesting fluxes of neutrinos from decay
of neutrons might be expected, e.g. in the case of neutrons produced in 
heavy nuclei photo-desintegration which do not lose efficiently 
energy in pion production in the radiation field.

In this review paper, we concentrate on the models of galactic neutrino 
sources which have been developed over the past few years. For
reviews on earlier works, the reader is referred to, e.g., 
Gaisser et al. (1995) and Learned \& Mannheim (2000).
In Section~\ref{sec:sn} we consider high energy neutrino 
fluxes produced during the early phase of supernova explosions,
whereas Section~\ref{sec:pwn} 
will be devoted to pulsar wind nebulae. 
Shell-type supernova remnants are also good candidate sources for high energy
neutrino emission, and this will be illustrated in Section 
\ref{sec:shell}. In Section~\ref{sec:pulsars} 
we will discuss neutrino emission from pulsars close to high density 
regions, such as those observed in the Galactic Center or massive stellar 
associations, and from neutron stars in binary systems in 
Section~\ref{sec:binary}. 
Finally, in Sections \ref{sec:mq} and \ref{sec:magnetars}
we will discuss models of microquasars and magnetars. 
Section \ref{sec:conc} will be devoted to our conclusions. 

The predicted event rates by the models reviewed in this paper
should vary in the presence of neutrino oscillations.
The expected flavor ratios at sources of high energy cosmic neutrino
fluxes from $pp$ or $p\gamma$ collisions are
$\phi_{\nu_e}:\phi_{\nu_{\mu}}:\phi_{\nu_{\tau}}=1:2:<10^{-5}$.
Given the solar neutrino results (see for instance Fogli et al.~2004),
and particularly the recent results by SNO (Ahmed et al.~2004)
with best fit oscillation parameters $\Delta m^2 = 7.1 \cdot 10^{-5}$ 
eV$^2$ and $\theta = 32.5$, the
strong evidence for $\nu_{\mu}$ and $\nu_{\tau}$ maximal
mixing from atmospheric neutrino experiments which prefer $\Delta m^2
\sim 2 \cdot 10^{-3}$ eV$^2$
(Super-Kamiokande (Fukuda et al.~1998),
MACRO (Ambrosio et al.~1998) and Soudan 2 (Sanchez et al.~2003)),
and the limit on the mixing matrix element 
$|U_{e3}|\ll 1$ from CHOOZ (Apollonio et al.~2003), these flavor 
ratios are
expected to transform into $1:1:1$ (Learned \& Pakvasa~1995, Athar et 
al.~2000) along $\gtrsim kpc$ distances in a direct hierarchy scenario.
For models considering electron antineutrinos from 
neutron decay (such as Anchordoqui et al.~2003a) described in 
Sec.~\ref{sec:massive}), about 60\% of the electron antineutrinos should
reach the Earth and $\sim 40\%$ should oscillate into $\bar{\nu}_{\mu}$
and $\bar{\nu}_{\tau}$ in about the same percentage.

In this paper we give an estimate of the muon neutrino event rates
in the absence of neutrino oscillations. 
If neutrino oscillations are to
be considered, a part of the muon neutrinos from mesons decay convert into
other flavors and could be detected. Nevertheless, the estimate of event rates
for $\tau$ events in neutrino telescopes needs a calculation including the
regeneration effect in the Earth due to the fast tau decay
(see Bugaev et al. 2004 and references therein). 
Moreover, it should be considered that the topologies induced by
$\nu_{\tau}$ and $\nu_e$ are different than $\nu_{\mu}$ induced events
and hence have different detection efficiencies.
As a matter of fact,
neutrino telescopes measure neutrinos indirectly thanks to their
interactions with the matter inside or surrounding instrumented regions
(for a review on recent results see Montaruli~2003).
The new generation of neutrino telescopes aims at the detection of
the Cherenkov light produced by charged secondaries induced in neutrino 
interactions. 3-D arrays of optical modules, 
pressure resistant glass spheres containing phototubes (PMTs),
are deployed in ice or sea/lake water depths. 
For muon neutrinos (and also for extremely high energy $\nu_{\tau}$),
the effective target mass is much larger than
the instrumented region, since the technique profits of the increase of the
muon (tau) range with energy. On the other hand, neutrinos of other flavors 
and neutral current interactions produce showers in the instrumented
regions.

\section{Event rates and luminosities from galactic sources}
\label{sec1}
For an isotropic emission from a source at a distance $D$, 
the relation between the luminosity and the neutrino energy flux 
$f_{\nu}$ is:
\begin{equation}
L_{\nu} =  4 \pi D^2 f_{\nu} \,.
\end{equation}
\noindent
A source of energy flux $f_{\nu}$ in neutrinos of energy $E_\nu$ will
produce N neutrino induced upward-going muons 
in a detector with effective area A and exposure time T: 
\begin{equation}
N = \frac{f_{\nu}}{E_{\nu}}P_{\nu \rightarrow \mu}  A T \, ,
\end{equation} 
\noindent
where the probability  $P_{\nu \rightarrow \mu}$ that a neutrino produces a muon above a certain
threshold depends on the neutrino charged current (CC) interaction 
cross section,
the muon range and should account for the Earth shadowing effect 
~\footnote{Earth shadowing effects are small for galactic models that 
predict neutrino emissions up to about 100 TeV. As a matter of fact the
neutrino interaction length is equal to the Earth diameter at about 40 TeV
(Gandhi~2000).}.
At 100 TeV the probability that a neutrino produces a muon of energy 
larger than 10 GeV is $P_{\nu \rightarrow \mu} \sim 10^{-4}$. 
Hence, the neutrino energy flux and the luminosity required to produce 
N detectable events are
\begin{equation}
f_{\nu} \sim  5 \cdot 10^{-12} N 
\left( \frac{E_{\nu}}{100 {\rm TeV}} \right)
\left( \frac{A T }{{\rm km^2 yr}}\right)^{-1}~~{\rm erg~\, cm^{-2} s^{-1}}
\end{equation} 
\noindent
and
\begin{equation}
L_{\nu} \sim  6 \cdot 10^{32} N
\left( \frac{D}{\rm kpc}\right)^{2}\times 
\left( \frac{E_{\nu}}{100 {\rm TeV}} \right)
\left( \frac{A T}{{\rm km^2 yr}}\right)^{-1} {\rm erg~\, s^{-1}}\,.
\end{equation} 
\noindent
This means that in one year the detection of 10 events of 100 TeV 
neutrinos coming from a source at D = 5 kpc in a 0.1 km$^2$ volume,
requires a luminosity of the order of $L_{\nu} \sim 10^{36}$ erg/s.
This large value of the luminosity strongly limits galactic source
candidates.

The reader should notice that a rigorous calculation of event rates 
requires integrals 
since the neutrino flux, the effective area and the cross sections  
are functions of the energy. 
A useful parameter to calculate event rates is the effective area 
for $\nu$'s, $A^{eff}_{\nu}$ (e.g. see Montaruli~2003, Ahrens et al.~2004a), 
that is the sensitive area
`seen' by $\nu$'s producing detectable $\mu$'s
when entering the Earth. In fact, the event rate 
in the detector from a point-like source at a declination $\delta$ 
producing a differential flux of neutrinos $\rm \frac{d\Phi}{dE_{\nu}}$, is
given by:
\begin{equation}
N_{\mu}(\delta) = 
\int_{E_{min}}^{E_{max}} dE_{\nu} A^{eff}_{\nu}(E_{\nu}, \delta) 
\frac{d\Phi}{dE_{\nu}} \, ,
\label{eq:evrates}
\end{equation}
where $E_{min}$ and $E_{max}$ are the minimum and maximum energies
of neutrinos for the considered flux, respectively.
$A^{eff}_{\nu}$ is given by:
\begin{equation}
A^{eff}_{\nu}(E_{\nu}, \delta) = \epsilon \cdot V_{gen} \cdot
N_{A} \rho \sigma_{\nu}(E_{\nu}) \cdot P_{Earth}(E_{\nu}) 
\label{eq:aeff}
\end{equation}
where $\epsilon = \frac{N_{sel}(E_{\nu},\delta)}{N_{gen}(E_{\nu},\delta)}$. 
$N_{gen}$ is the number
of generated events in the generation volume $V_{gen}$ (whose 
dimensions depend on the muon range) and $N_{sel}$ is the number of
selected events that depends on track reconstruction quality cuts 
and selection criteria for background rejection; 
$\rho \cdot N_A$ is the number of target nucleons per unit
volume, with $N_A$ the Avogadro number and $\rho$ the material density,
$\sigma_\nu$ is the neutrino CC cross section and 
\begin{equation}
P_{Earth}(E_\nu) =
e^{- N_A \sigma(E_{\nu}) \int \rho(l) dl}
\end{equation}
\noindent
is the neutrino absorption 
probability in the Earth for a given neutrino of energy $E_{\nu}$ that 
propagates along paths of density $\rho(l)$. The path in
the Earth depends on the neutrino direction.

The rejection of the residual atmospheric muons reaching detector
depths can be achieved selecting upward-going muons induced by muon
neutrinos interacting in the surrounding material. 
The rejection of backgrounds due to atmospheric neutrinos can be achieved 
through angular cuts around candidate source directions, 
which mainly depend on the experimental angular resolution. 
For a telescope with sub-degree accuracy the number of atmospheric neutrinos 
is $\lesssim 1$ event/yr/km$^2$. 
Also energy cuts can be applied to reduce further the
backgrounds. Nevertheless, for galactic source models
predicting neutrino emissions in the $1-100$ TeV energy range, 
the tail of the atmospheric neutrino spectrum
can still dominate over cosmic neutrino spectra. Hence angular cuts and
possibly coincidence requirements with pulsed emissions or time modulations
are more effective.

Experimental results on neutrino searches from discrete galactic
sources have been obtained by detectors of area $< 1000$ m$^2$ which have
produced the first limits: IMB (Svoboda et al.~1987; Becker-Szendy et al.~1995),
Kamiokande (Oyama et al.~1989) and Kolar Gold Fields (Adarkar et al.~1991). 
Subsequently, limits were presented by the MACRO experiment
(Ambrosio et al.~2001, updated in Montaruli~2003). Recently also
Super-Kamiokande presented results for discrete steady sources 
(Washburn et al.~2003).

The new generation neutrino telescope AMANDA in the South Pole ice, 
with area larger than $10^4$ m$^2$, has not yet detected any excess from 
discrete sources, although it has set interesting  
limits for constraining models (Ahrens et al.~2003; Ahrens et al.~2004a).
AMANDA-II sensitivity, close to what expected 
for the ANTARES Mediterranean undersea neutrino telescope 
(Amram et al.~1999, Heijboer et al.~2003), could be sufficient to detect a signal 
among the atmospheric neutrino background for some of the models
that are presented in this review (e.g. those in Sec.~\ref{sec:sn}
and Sec.~\ref{sec:mq}). 
Most of the models considered in this review predict neutrino event rates 
above the sensitivity limit of the future cubic-kilometer size detector
IceCube (Ahrens et al.~2004b) and of a possible km$^3$ detector in the 
Mediterranean, as proposed to the European Community (KM3NeT~2004) 
and by the NEMO collaboration (Piattelli et al.~2003).

\begin{figure}[htb]
  \begin{center}
    \includegraphics[height=20.pc,width=20.pc]{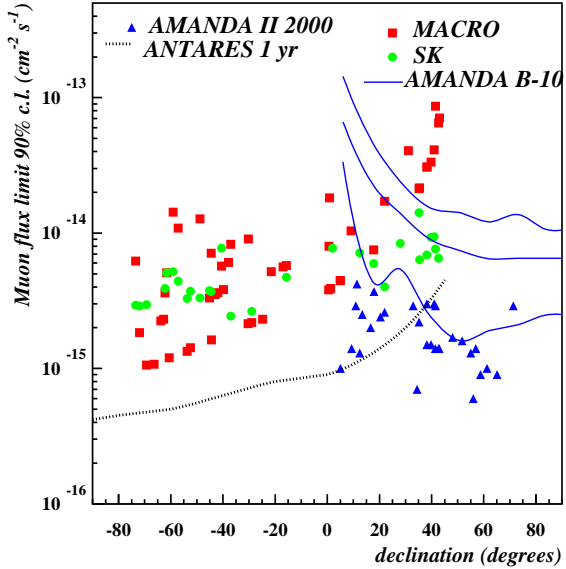} 
  \end{center}
  \caption{90\% c.l. upper limits on $\mu$ fluxes induced 
by $\nu$'s with $E^{-2}$ spectrum vs source
declination for SK (green circles, Washburn et al.~2003), 
MACRO (red squares, Montaruli~2003, previous results in 
Ambrosio et al.~2001), AMANDA-B10 (3 blue solid lines, Ahrens et al.~2003), 
AMANDA-II 2000 data (blue triangles, Ahrens et al.~2004a),
ANTARES sensitivity after 1 yr (dotted black line, Heijboer et al.~2003).
It has not been possible to apply a correction due to different
$\mu$ average energy thresholds (SK $\sim 3$~GeV, MACRO
$\sim 1.5$ GeV, AMANDA $\sim 50$~GeV). Nevertheless, the 
maximum of the response curves of these
detectors for an $E^{-2}$ flux is at $E_{\mu} \sim 10$~TeV, hence 
events contributing between 1-50 GeV should not
make a large correction to these limits. 
}
\label{fig1}
\end{figure}

Recent results on upper limits on the neutrino induced $\mu$ fluxes 
as a function of declination for discrete sources are shown in 
Fig.~\ref{fig1} (Montaruli~2003).
In the region of declination observed by South Pole experiments,
triangles indicate the upper limits obtained
by AMANDA II (Ahrens et al.~2004a)  using 699 upward events 
selected in 197 d during the year 2000. The most significant excess, observed 
around 21.1h R.A. and $68^{\circ}$ declination, 
is of 8 events and the expected background is 2.1. The probability 
to observe such an excess as a random fluctuation 
of the background is 51\%. Three solid lines indicate the limits
by AMANDA B-10 obtained in 130 days
(Tab.~4 in Ahrens et al.~2003), giving an idea
of the effect of the statistical fluctuations of limits that are
given in declination bands for different right ascension bins. The 
lower line connects the lowest limits obtained in a declination band 
and the upper line connects the highest ones. The middle line connects the 
average of the limits in right ascension bins for each
declination band. 
Limits from the experiments in the upper hemisphere are shown:
Super-Kamiokande (Washburn et al.~2003, circles) and MACRO 
(Montaruli~2003, squares).
Super-Kamiokande, with an angular resolution of about $2^{\circ}$,
reported results using a sample of 2369 upward-going muons collected in 4.6
yrs.
The MACRO scintillator+tracking detector, with angular resolution 
$\lesssim 1^{\circ}$, reported results using
a sample of 1388 upward-going $\mu$'s collected in 6 yrs(Montaruli~2003).
Ten neutrino events were measured inside a $3^{\circ}$ half-width cone
(that is expected to include $90\%$ of an $E^{-2}$ neutrino signal) 
around the plerion PSR1509-58, while
2 are expected from atmospheric $\nu$'s. Even though PSR1509-58 is a source
of interest as possible $\nu$ emitter (see Sec.~\ref{sec:pwn}), 
the significance of MACRO result is negligible 
when all the 1388 directions of the measured events are looked at. 
Moreover, it is
expected that $E^{-(2 \div 2.5)}$ signals should produce at least 4-7 events 
in $1.5^{\circ}$ around the source while only 1 is detected and the expected
background is 0.5.  
A $\gamma$ emission above 1.9 TeV was observed by CANGAROO from this source
in 1997 with $4\sigma$ significance, but not confirmed by the 1996 and 1998
data analysis with 2.5 TeV threshold (Sako et al.~2000).
Models concerning
this source are described in Sec.~\ref{sec:pwn} and predict much lower
rates than what would be needed to explain the observed
events in MACRO.
From the same sources Super-Kamiokande (A. Habig, private communication) 
observes the largest number of
events between the selected catalogue looked at, but the data are still
compatible with background fluctuations
(9 events to be compared to a background of 5.4).
The sensitivity of ANTARES 
(expected angular resolution of $\sim 0.2^{\circ}$ for $E_{\nu}>10$ TeV) 
is also shown for 1 yr of data taking (Heijboer et al.~2003). 

\section{Early phase of supernova explosion}
\label{sec:sn}

The supernovae can provide promising event rates during the 
relatively short period after the explosion. Nevertheless
it should be considered that they are
rare events, since the expected supernova rate in the Galaxy
is of the order of 2-4 per century and the pulsar birth rate is
one per 20 -250 yrs (Lorimer~2003, Vranesevic et al.~2003). 
\subsection{Supernova shock waves}
\label{sec:sn1}

Type II supernovae are created by massive stars with extended envelopes.
When the supernova shock breaks out of such a star, it creates a 
collisionless shock which may accelerate protons above 10 TeV 
(Waxman \& Loeb 2001). If the
protons take a significant fraction of the post-shock energy, then 
observable fluxes of TeV neutrinos lasting $\sim$ 1 hr can be produced 
in the inelastic collisions 
of relativistic protons with matter, about 10 hrs after the
initial low energy neutrino flash following the core collapse of the 
star. The model predicts about 100 events in a 1 km$^2$ detector caused by
muon neutrinos from a supernova at a distance of 10 kpc,
if the proton acceleration efficiency is close to 10$\%$.
Clear predictions concerning the time lag between the low and high energy 
neutrino emission
should allow to distinguish between the proposed scenario and other models of
neutrino production during the early phase of the supernova explosion.  
\subsection{Supernovae with energetic pulsars}
\label{sec:snpulsar}

Young supernova remnants (SNRs) have been proposed as sites of particle
acceleration to high energy and, as a consequence, also possible sources of
high-energy neutrino and $\gamma$-ray emissions. 
The first calculations of the neutrino flux from SNRs
containing energetic pulsars have been performed by 
Sato~(1977), Berezinsky \& Prilutsky (1978), 
and more recently, also in relation to the explosion of {\it SN 1987A}
in the Large Magellanic Cloud (LMC), e.g. by
Gaisser et al.~(1987), Gaisser \& Stanev~(1987), 
Berezinsky \& Ginzburg~(1987),
 Berezinsky et al.~(1988),
Yamada et al.~(1988), and Gaisser et al.~(1989).

In this review, we consider more recent models, starting from the one
developed by Protheroe et al. (1998). In this model the authors 
suppose that a pulsar with sufficiently short rotational 
period and high magnetic field has been formed just after the supernova 
explosion. Heavy ions, mainly $^{56}Fe$ nuclei, can be accelerated from 
the star surface along open magnetic field lines into the pulsar 
magnetosphere. The slot gap model (see Arons \& Scharlemann 1979) for the 
acceleration of nuclei has been applied. 
$^{56}Fe$ nuclei, extracted from the star surface by thermoionic emission, 
are gradually accelerated in the slot gap and suffer
photodisintegrations in collisions with the thermal photons 
coming from the neutron star surface.
To check if photodisintegration of $^{56}Fe$ nuclei is important, 
Protheroe {et al. (1998) compute the average number of neutrons 
extracted from $^{56}Fe$ nuclei as a function of their Lorentz factor,
assuming that in each photodisintegration only one neutron is 
extracted. It is found that this process occurs efficiently only
if the surface temperature is quite high, above $\approx 10^7 K$,
i.e. within a few days following the explosion. 
Some neutrons can be still extracted
from the nuclei during the first year after the explosion, 
when the surface temperature drops 
to $\sim 4\times 10^6$ K (Nomoto \& Tsuruta~1987).
However, neutron star polar caps may have high temperatures, thus
leading to efficient photodisintegration of iron nuclei emitted from
the hot polar caps. 
The value of the temperature turns out to be dependent 
on the value of the magnetic field, the initial pulsar period
and its radius. 

Relativistic neutrons extracted from Fe nuclei in this way move ballistically
through the magnetosphere and beyond the light cylinder. Their Lorentz factors,
which lie within the range $\sim 10^4-10^7$,
depend on the pulsar parameters and on the temperature of the 
radiation. Typical values of the temperature are of the order of
$\sim 10^6-10^7~ \rm K$
for periods shorter than $\sim 10$ ms and surface magnetic fields 
$10^{12}$ G. Such neutrons decay at a distance of $\sim 0.1-100$ pc from the 
pulsar. Initially, when the SNR shell is still opaque to neutrons, they 
mostly interact with target nuclei as they travel out through the dense 
SNR shell, thus producing $\gamma$-rays and neutrino signals. 

Protheroe et al. calculate the ($\nu_\mu + \bar{\nu}_\mu$) neutrino spectra 
for a  source 
located  at a distance $\rm d=10~kpc$, at a time $\rm t=0.1~year$ after 
the explosion, assuming initial periods of the pulsar $P_0=5~\rm ms$ 
and $P_0=10~\rm ms$.
The magnetic field is kept fixed at $\rm B=10^{12}~ G$.
Since the ion emission is dependent on the temperature of the star surface
and the polar cap, two limiting cases have been considered, i.e.
\begin{itemize}
\item[i)] "no polar cap heating", where the whole star has temperature 
equal to the surface temperature;
\item[ii)] "maximum polar cap heating", where the polar cap has temperature 
larger than the surface temperature.
\end{itemize}
In the following, we report the neutrino energy spectra, calculated 
in units of $\rm GeV^{-1}~m^{-2}~s^{-1}$, for an initial pulsar period
$P_{0}\rm = 10~ ms$ (see Fig. 9 in Protheroe {\it et al.} 1998) 
\begin{eqnarray}
\frac{dN_\nu}{dE_\nu}&=&3.220\times10^{-5}\,
\left(\frac{E_\nu}{1 \mbox{GeV}}\right)^{-1.680}~~(\rm E_\nu<10^4~GeV) \nonumber\\
         &=&2.387\times10^{3}\,
\left(\frac{E_\nu}{1 \mbox{GeV}}\right)^{-3.622}~(\rm E_\nu>10^4~GeV)~(no~h.) \nonumber \\
\label {10ms} 
\end{eqnarray} 
\begin{eqnarray}
\frac{dN_\nu}{dE_\nu}&=&1.058\times10^{-4}\,
\left(\frac{E_\nu}{1 \mbox{GeV}}\right)^{-1.800}~~(\rm E_\nu<10^4~GeV) \nonumber\\                 
&=&2.387\times10^{3}\,
\left(\frac{E_\nu}{1 \mbox{GeV}}\right)^{-3.622}~(\rm E_\nu>10^4~GeV)~(max.~ h.) \nonumber \\  
 \label{10ms_max}
\end{eqnarray} 
\noindent
For an initial period $P_{0}\rm = 5~ ms$, the neutrino spectra are:
\begin{eqnarray}
\frac{dN_\nu}{dE_\nu}&=&4.470\times10^{-5}\,
\left(\frac{E_\nu}{1 \mbox{GeV}}\right)^{-1.480}~~(\rm E_\nu<10^4~GeV) \nonumber\\
         &=&1.126\times10^{6}\,
\left(\frac{E_\nu}{1 \mbox{GeV}}\right)^{-3.687}~(\rm E_\nu>10^4~GeV)~(no~h.) \nonumber \\
\label {5ms} 
\end{eqnarray} 
\begin{eqnarray}
\frac{dN_\nu}{dE_\nu}&=&9.120\times10^{-4}\,
\left(\frac{E_\nu}{1 \mbox{GeV}}\right)^{-1.788}~~(\rm E_\nu<10^4~GeV) \nonumber\\                 
&=&1.126\times10^{6}\,
\left(\frac{E_\nu}{1 \mbox{GeV}}\right)^{-3.687}~(\rm E_\nu>10^4~GeV)~(max.~ h.) \nonumber \\  
 \label{5ms_max}
\end{eqnarray} 

\noindent
From these calculations it is clear that selecting an angular region of 
$1^{\circ}$ around the source, the neutrino flux above 1 TeV is well above the 
atmospheric neutrino background.
For a $10^{\circ}$ region, a neutrino signal from a galactic source 
for the $P_0=5~\rm ms$ case
may be observable above the background at energies above 10 TeV. 
However, as shown in Protheroe et al. (1998), the neutrino light curve drops 
rapidly at $\rm t > 0.1~ year$, and the signal could be detected among the
atmospheric background only in detectors of dimensions
of the order of km$^2$.

Since the neutrino emission from acceleration and subsequent fragmentation 
of nuclei is beamed, the neutrino signal should be
strongly anisotropic and modulated with the pulsar period.
Hence the signal could be observed only if the beam is in the
Earth direction.

Another target for hadrons accelerated by a pulsar early after the supernova 
explosion can be provided by the thermal emission of the expanding opaque
supernova envelope. Beall \& Bednarek (2002) consider neutrino production 
in collisions of nuclei with the radiation filling the
cavity below the hot expanding supernova remnant envelope. The heavy 
nuclei, accelerated close to the light cylinder of the 
pulsar or in the pulsar wind region, are completely disintegrated in 
collisions with the dense radiation field at early times after the
supernova explosion. 
The resulting products (protons and neutrons) 
lose energy efficiently in pion production. 
However, due to the very dense radiation
field, pions can also lose energy before decaying due to 
Inverse Compton Scattering (ICS).
These ICS energy losses of pions are important during the first 
$\sim 10^4$ s after 
supernova explosion. However, later pions decay into muon neutrinos which
escape without absorption through the expanding supernova envelope.
When the column density of matter through the envelope drops to $\sim 10^3$ 
g cm$^{-3}$, the radiation is not further confined by the envelope and the
nuclei interact with the envelope matter producing muon neutrinos. 
Muon neutrino fluxes have been calculated from interactions of nuclei with
the radiation field inside the envelope during the period $10^4 \div 2
\cdot 10^6$ s after the supernova explosion and from interactions 
with the matter envelope during the period $2 \cdot 10^6 \div 3 \cdot 10^7$
s after the explosion. During these periods the density is low enough 
so that pions produced in the interactions can decay. At later times
the neutrino flux drops significantly because the column density of matter 
through the envelope drops to several g cm$^{-3}$ and particles are not
completely cooled in collisions, hence they do not lose
energy efficiently.

The model predicts up to $\sim 10^4-10^5$ events in a 1 km$^2$ 
neutrino detector from a supernova at a distance of 10 kpc during the
first year after the explosion, for pulsars with initial period of the
order of a few milliseconds and typical surface magnetic fields of a few 
$10^{12}$ G, or pulsars rotating with a period of 10 ms and with surface 
magnetic field of the order of $10^{14}$ G provided that the acceleration 
efficiency of nuclei by the 
pulsar is $\xi = 1$, i.e. all rotational energy of 
the pulsar goes into nuclei acceleration. 
Hence, these results hold if these pulsars are able to accelerate nuclei up to
energies of the order of $10^{20}$ eV.
The event rates, obtained 
by using the neutrino detection probability in Gaisser \& Grillo 
(1987) and the absorption coefficients of neutrinos in the Earth calculated
by Gandhi (2000), are given in Tab.~\ref{tab1} of Beall and Bednarek (2002)
for $\xi = 1$. 
The neutrinos are produced with a $E_\nu^{-1}$ power law differential 
spectrum between $\sim 0.1 - 100$ PeV 
for the case of hadron interactions with radiation,
and up to $\sim 10$ PeV for hadron interactions with the envelope matter
(see Figs. 2 and 3 in Beall \& Bednarek 2002). 

However, the value of the acceleration efficiency $\xi$ is a free parameter 
of the model, and might be
constrained by observations. For this purpose, Blasi et al. (2000) propose
that heavy nuclei injected by fast pulsars into the Galaxy, after the 
supernova becomes transparent, can explain the observed cosmic ray (CR)
flux at 
$\sim 10^{20}$ eV. In fact, from the normalization of the flux of nuclei 
predicted in this model
to the observed cosmic ray flux in the Galaxy (see Eq.~8 in Blasi et 
al. 2000), the acceleration efficiency of a pulsar can be estimated, 
and the following relation can be obtained
\begin{eqnarray}
\xi \epsilon Q/(\tau_2 R_1^2 B_{13})\approx 4\times 10^{-6},
\label{eq:eff}
\end{eqnarray}
\noindent
where $\epsilon$ is the fraction of pulsars 
able to accelerate iron nuclei to
$10^{20}$ eV, $Q$ is the trapping factor of nuclei in the Galactic halo, 
$\tau = 100\tau_2$ is the rate of neutron star formation, $R = 10 R_1$ kpc 
is the radius of 
the Galactic halo, and $B = 10^{13}B_{13}$ G is the surface magnetic field 
of the pulsar.
Applying reasonable values to these parameters, 
$Q = 1$, $\tau_2 = 1$, $R_1 = 3$, 
$\epsilon = 0.1$, and $B_{13}$ also considered in the pulsar model,
the estimated efficiencies of pulsar acceleration $\xi$ 
are $3.6 \cdot 10^{-3}$ for $B_{12} = 100$ and 
$1.4 \cdot 10^{-4}$ for $B_{12} = 4$. 
After taking into account these values of $\xi$ 
the event rates are shown in the first two columns of Tab.~\ref{tab1} for the
indicated pulsar parameters.  
The muon neutrino event rates expected for horizontal and vertical directions, 
after including the absorption by the Earth, are indicated by (H) and (V), 
respectively. 
The reactions N-$\gamma \rightarrow \nu_\mu$ and Fe-M$\rightarrow \nu_\mu$ 
are used for the process of neutrino production in hadron (nucleons or iron 
nuclei) collisions with the radiation 
and with the envelope matter, respectively.
The model predicts up to hundreds of neutrino events in a 
1 km$^2$ detector for pulsars, with initial 
periods of several ms and surface magnetic fields 
of the order of a $10^{14}$ G,at a distance of 10 kpc 
provided that $\xi > 10^{-3}$. 
On the other hand, Crab-like pulsars at a distance of 10 kpc, 
with initial periods of several ms and surface magnetic fields 
of the order of a few $10^{12}$ G, can produce detectable event rates in a 
1 km$^2$ detector provided that $\xi$ is not far from 1 (see third column
in Tab.~\ref{tab1}). 
As a matter of fact, a pulsar with a period of 20 ms and a magnetic field
of $4 \cdot 10^{12}$ G cannot accelerate nuclei up to $10^{20}$ eV, hence 
the acceleration efficiency is not constrained by eq.~\ref{eq:eff}.
Hence, as an example, event rates are given in the optimistic
assumption of fully efficient pulsar acceleration.

\begin{table} 
      \caption{($\nu_\mu + \bar{\nu}_{\mu}$) induced events from a supernova
at distance of 10 kpc in the horizontal (H) and vertical (V) directions
in a km$^2$ detector. The calculation is done for different
pulsar periods and magnetic fields (from Tab.~1 in Beall and Bednarek
(2002)). The calculation of rates includes the indicated values
of the acceleration efficiency of the pulsar $\xi$ (see eq.~\ref{eq:eff}). 
} 
\begin{center}
	 \begin{tabular}{|c|c|c|c|} 
	    \hline 
	     ~pulsar~~$\Rightarrow$
& $P_{\rm ms} = 3$ & $P_{\rm ms} = 10$ & $P_{\rm ms} = 20$\\ 
	     process $\Downarrow$
& $B_{12} = 4$     & $B_{12} = 100$    & $B_{12} = 4$\\ 
	     \hline 
	    N-$\gamma \rightarrow \nu_\mu$ (H)&  1.2& 
86.4 & ~~~2.6\\   	    
            \hline 
	    Fe-M$\rightarrow \nu_\mu$ (H)      & 12.2 & 
648.0 & ~~11.3\\
 	    \hline
	    N-$\gamma \rightarrow \nu_\mu$ (V)& 0.3  & 
18.4  & ~~~1.1\\
   	    \hline 
	    Fe-M$\rightarrow \nu_\mu$ (V)      & 4.8  & 
230.4 & ~~~8.6\\
 	    \hline 
	    $\xi$     & $1.4\times 10^{-4}$  &  $3.6\times 10^{-3}$ & 1\\
 	    \hline 
	 \end{tabular} 
\end{center}
	 \label{tab1} 
     \end{table} 

Neutrino production in high energy hadron-hadron collisions in
young pulsar driven supernovae has been further investigated by Nagataki~(2004). 
It is assumed that most of the rotational energy of a pulsar with millisecond 
initial period is taken by relativistic hadrons with the Lorentz factor 
kept as a free parameter of the model. Hadrons, accelerated in the pulsar 
wind region, are thermalized in the inner shock and interact inside the 
pulsar wind nebula between themselves. 
The neutrino event rates with energies above 10 GeV expected 
from a 1 ms pulsar at a distance of 10 kpc, are above the atmospheric neutrino background, 
and could be detected by cubic-kilometer neutrino telescopes during the first
100 yrs after the explosion. However, note that according to this modeling the shock 
in the pulsar wind is located very close to the pulsar, at a distance of 
$\sim 10^{12}$ 
cm (for a pulsar with initial period 1 ms, surface magnetic 
field $10^{12}$ G, and age of 100 yrs). Hence, the density of relativistic hadrons 
can be high enough for efficient production of pions, which subsequently decay into 
$\gamma$-rays and neutrinos. Nevertheless, it seems that this model may 
have difficulties in predicting the observed location of the shock in the Crab Nebula, 
which is at a distance of 
$\sim 3\times 10^{17}$ cm at its present age of $\sim 10^3$ yrs. 
Unless this constraint is fulfilled, the predicted fluxes of neutrinos 
could be not reliable, 
since, as the author shows in the paper, the expected neutrino flux is very sensitive 
to the shock location. Moreover, the amount of energy supplied by the 
pulsar to the expanding nebula is not taken into account. 
For a pulsar with 1 ms initial period and  magnetic field $10^{12}$ G, 
the energy of relativistic protons, which is equal to the rotational energy 
lost by the pulsar during the first 10 yrs ($\sim 10^{52}$ ergs), is much 
larger than the initial 
kinetic energy of a typical supernova envelope ($\sim 10^{50}$ ergs for the Crab Nebula).
This additional source of energy should accelerate the nebula, moving the location of the
shock outwards. As a result, the relativistic hadron densities decrease
and the expected neutrino fluxes should be much smaller.  

\section{Pulsar wind nebulae (Plerions)}
\label{sec:pwn}

A number of plerions have been discovered in radio, optical, and
X-ray bands, with the Crab as the youngest and most energetic source.
Only about 10\% of supernova remnants are classified as plerions. 
Plerionic nature is usually indicated by a center-filled morphology,
resulting from the continuous injection of pulsar electrons into the 
nebula, with the additional constraint that the spectra must be non-thermal
(but power-laws) resulting from a statistical acceleration process.
The radio, optical and X-ray emission observed from plerions is believed to be
due to synchrotron emission. ICS is usually 
seen at higher energies compared to synchrotron emission. 
It is likely that hadrons can also play an important role
in radiation processes (e.g. Cheng et al. 1990, Aharonian \& Atoyan~1995).

In order to be more quantitative, let us suppose that TeV photons produced 
by a plerion are the decay products of $\pi^0$'s. Neutral pions can be
produced by accelerated protons in either $pp$ or $p\gamma$ collisions,
depending on the relative target density of photons and protons in the source
region where protons are accelerated. The ($\nu_\mu + \bar \nu_\mu$) 
neutrino flux, $\rm dN_\nu/dE_\nu$, produced by the decay of charged pions
in the source can be derived from the observed $\gamma$-ray flux,
$\rm dN_\gamma/dE_\gamma$, by imposing energy conservation and neglecting
photon absorption:
\begin{equation}
\int_{E_\gamma^{min}}^{E_\gamma^{max}} E_\gamma \frac{dN_\gamma}{dE_\gamma}
dE_\gamma = K \int_{E_\nu^{min}}^{E_\nu^{max}} E_\nu \frac{dN_\nu}{dE_\nu}
dE_\nu \label{scale}
\end{equation}
\noindent
where $E_\gamma^{min}$ ($E_\gamma^{max}$) is the minimum (maximum) energy of 
the photons that have hadronic origin.
$E_\nu^{min}$ and $E_\nu^{max}$ are the corresponding minimum  and
maximum energies of the neutrinos. The factor K depends on whether the
$\pi^0$'s are of $pp$ or $p\gamma$ origin. As a first order
approximation, using average values, it is generally assumed that 
$\rm K=1$ for $pp$ scattering, whereas $\rm K=4$ for $p\gamma$ interactions.

Such simple rescaling of the $\gamma$-ray fluxes to the neutrino fluxes
has been done by Guetta and Amato (2003) for a few pulsar wind nebulae
(PWNe) observed in
TeV $\gamma$-rays, i.e. the Crab
Nebula, the Vela SNR, the pulsar wind nebula around PSR1706-44 and
the nebula surrounding PSR1509-58. 
For instance, the neutrino flux predicted for the Crab nebula
(Guetta 2003, private communication) is given 
(in units of $\rm GeV^{-1}~m^{-2}~s^{-1}$) by: 
\begin{equation}
  \frac{dN_\nu}{dE_\nu}=1.85 \times10^{-10}\left(\frac{E_\nu}{10^3\mbox{GeV}}
  \right)^{-2.6} ~~(\rm E_\nu < 10^5~GeV) \, .
  \label{Crab}
\end{equation}    
\noindent
It is concluded that all these PWNe could be detected by a 1 km$^2$ 
neutrino detector with event rates per year 
between 1 up to about 12 in the energy range $E_\nu\sim 1 - 100$ TeV.
However, these estimations are based on the assumption 
that all $\gamma$-rays above 2 TeV are produced in $pp$ collisions via 
pion decay. It has been argued by Bednarek \& Bartosik (2003)
that this is a reasonable approximation for relatively young
nebulae, of the age of the Crab nebula or younger, but not for older 
nebulae, such as those mentioned above.

A more specific model for $\gamma$-ray and neutrino production in the Crab 
Nebula has been proposed a few years ago by Bednarek and
Protheroe (1997). The general scenario is similar to the one
previously discussed for the very young supernova remnants 
(Protheroe et al.~1998): 
iron nuclei from the Crab pulsar are accelerated in the outer gap 
(Cheng et al.~1986) and photodisintegrate in 
collisions with the non-thermal radiation field filling the outer gap,
thus producing energetic neutrons which are injected into the nebula.
Neutrons can decay either inside or outside the Crab nebula. Those neutrons
which decay inside the nebula produce protons, which collide with 
the matter in the nebula producing $\gamma$-rays and neutrinos.
Neutrons which decay outside the nebula can be captured by the expanding
nebula at later times or contribute to galactic cosmic rays. 
The neutrino fluxes predicted by this model are dependent
on the Crab pulsar and nebula parameters such as: the initial pulsar 
period, the nebula radius and the mass ejected during the  
supernova explosion. When reasonable values are taken into calculations, then
the expected neutrino spectrum (in units of $\rm GeV^{-1}~m^{-2}~s^{-1}$)
can be described by the following  broken power law: 
\begin{eqnarray}
  \frac{dN_\nu}{dE_\nu}& = & 1.735\times10^{-6}\,
\left(\frac{E_\nu}{1 \mbox{GeV}}\right)^{-1.61}~~(\rm E_\nu \lesssim 10^4~GeV) \nonumber\\
         & = &  1.47\times10^{2}\,
\left(\frac{E_\nu}{1 \mbox{GeV}}\right)^{-3.57}~~(\rm E_\nu \gtrsim 10^4~GeV) \, .
  \label{Crab_bis}
\end{eqnarray}    
\noindent
If the observed TeV $\gamma$-ray emission from the Crab 
Nebula is  due to the hadronic processes at energies above 10 TeV, then the 
model predicts
a few neutrino events per year in a 1 km$^2$ detector.

Recently, two papers discuss independently 
the hadronic processes in the PWNe,
assuming that a significant part of the rotational energy lost by the pulsar
is taken by relativistic heavy nuclei accelerated in the pulsar wind region
(Bednarek \& Bartosik 2003, Amato et al.~2003). 
This assumption can explain some fine structures in the Crab Nebula, 
as discussed by, e.g., Gallant \& Arons (1994). 
According to their model, 
the nuclei can generate Alfv\'en waves just above the pulsar wind shock,
which resonantly scatter off positrons and electrons, accelerating them to the
energies observed, e.g., in the Crab Nebula. Relativistic nuclei injected into 
the nebula can interact with the nebula matter, and produce 
$\gamma$-rays and neutrinos via pion decay. 
The consequence of this hybrid hadronic-leptonic model is that 
it self-consistently treats both leptonic 
and hadronic processes.

In the model proposed by Bednarek \& Bartosik (2003),
the basic parameters of the expanding nebula,
the injection spectra of hadrons and leptons, and their equilibrium
spectra inside the nebula, depend on time,
according to the expectations of the pulsar evolution model.
Some free parameters of the model are fixed by the comparison with the TeV 
$\gamma$-ray emissions from PWNe, such as the Crab Nebula, the Vela nebula,
and the nebula around PSR 1706-44. Based on that, they predict the TeV $\gamma$-ray 
fluxes from PWNe which contain energetic pulsars.
The correct modeling of the known X-ray emission from the TeV $\gamma$-ray 
PWNe is possible if it is assumed that the lepton acceleration efficiency 
by nuclei drops with the age of the nebula. This suggests that
the lepton acceleration efficiency for a specific nebula can also depend
on time. This effect has not been included into the calculations.
In this model, Bednarek (2003a) calculates the neutrino spectra from a few PWNe 
and predicts also the neutrino event rates
in a 1 km$^2$ detector from a few PWNe. 
It is concluded that the largest rate is from the Crab Nebula 
that would produce a rate larger than 1 event per year 
(using the detection probabilities by Gaisser \& Grillo 1987).
In this case, the predicted spectrum is above 
the atmospheric neutrino background in the energy range 
$\sim 10 - 100$ TeV, and is given (in units of $\rm GeV^{-1}~m^{-2}~s^{-1}$)  by:
\begin{eqnarray}
  \frac{dN_\nu}{dE_\nu}& = & 4.045\times10^{-8}\,
\left(\frac{E_\nu}{1 \mbox{GeV}}\right)^{-1.44}~~(\rm
E_\nu<6 \cdot 10^4~GeV) \nonumber\\
         & = &  3.14\times10^{14}\,
\left(\frac{E_\nu}{1 \mbox{GeV}}\right)^{-6.11}~~(\rm E_\nu>6
\cdot 10^4~GeV)
  \label{Crab_ter}
\end{eqnarray}    
Other PWNe can be detected only if they are close to high density
regions, which is probably the case of MSH15-52 around PSR1509-58.
The ratio of the TeV neutrino and  $\gamma$-ray fluxes depends on the 
age of the PWNa in this hybrid hadronic-leptonic model, dropping from
$\sim 58\%$ for the Crab Nebula, down to $22\%$ for the Vela nebula, and
$0.5\%$ for the nebula around PSR 1706-44.  

Amato et al. (2003) present an analytical approach to the leptonic and 
hadronic processes inside the PWNe. The neutrino event rates predicted by
them for the Crab Nebula in a 1 km$^2$ neutrino detector are between 4 
and 14, depending on the Lorentz factor of nuclei injected by the pulsar.
They are significantly higher than those obtained by Bednarek (2003a). 
The discrepancies between the models might be ascribed, on one hand, to the 
different treatment of the injection spectra of hadrons, which actually 
evolve in time. On the other hand, the neutrino spectra are determined by 
the hadronic equilibrium spectra inside the nebula, and these are constrained 
by the observations. A time dependent treatment 
of the equilibrium spectra is required in order to have the correct
neutrino spectrum. In the model by Amato et al. (2003), the Lorentz 
factors of the nuclei are assumed to be constant and independent on 
the pulsar parameters, which actually should evolve in time due to the pulsar 
rotational energy losses.

\section{Shell-type supernova remnants}
\label{sec:shell}

A few spherical supernova remnants, such as 
SN1006 (Tanimori et al.~1998), Cas A (Aharonian et al.~2001), and  
SNR RX J1713.7-3946 (Enomoto et al.~2002),
have been reported to emit $\gamma$-rays with energies above $\sim 1$ TeV. 
From these observations, it cannot be established whether 
the mechanism of $\gamma$-ray production in these sources is leptonic or 
hadronic. In other words, these high-energy photons could be produced either
by electron beams via ICS (Pohl~2001), or by 
proton beams via the production and subsequent decay of neutral pions
(Gaisser et al.~1995; Pohl~2001). In the latter case, the high-energy photons
are accompanied by neutrinos because charged pions are produced along with 
$\pi^0$'s.

Recently the CANGAROO collaboration has shown that the energy spectrum of the 
$\gamma$-ray emission from the supernova remnant RX J1713.7-3946
matches that expected if the $\gamma$-ray are the products of $\pi^0$ 
decay generated in $pp$ collisions (Enomoto et al.~2002). 
Alvarez-Muniz \& Halzen (2002) have estimated the 
expected rate of neutrino events in a 1 km$^2$ detector
using the normalization between
$\gamma$-ray and neutrino fluxes (see eq.~\ref{scale}). 
Assuming that all charged pions decay in the environment of the supernova and
that the neutrino spectrum follows the input proton spectrum, assumed to be of 
$\sim E^{-2}$ type, the calculated ($\nu_\mu + \bar\nu_\mu$) flux is given
(in units of $\rm GeV^{-1}~m^{-2}~s^{-1}$) by: 
\begin{equation}
  \frac{dN_\nu}{dE_\nu} =  4.14\times10^{-4}\, \rm 
\left(\frac{E_\nu}{1 \mbox{GeV}}\right)^{-2}~~~~(\rm E_\nu < 10^4~GeV)
  \label{RX}
\end{equation}    
\noindent
Alvarez-Muniz \& Halzen have predicted large muon neutrino
event rates from SNR RX J1713.7-3946, of the order of 
$\sim 40$ km$^{-2}$ yr$^{-1}$. However, the reader should notice that 
the hadronic origin of $\gamma$-rays 
from this supernova remnant has not been confirmed by the analysis of
hadronic and leptonic processes performed by 
Reimer \& Pohl~(2002).

Neutrino fluxes have been predicted by Alvarez-Muniz \& Halzen
for some other supernovae, such as, Cassiopeia A, Crab Nebula,
and the shell-type supernova remnant Sagittarius A East at the Galactic Centre 
(Buckley et al.~1997). For the SNR Cassiopeia A, the predicted 
neutrino event rate is $\sim~5~\rm km^{-2}~ yr^{-1}$, whereas 
for Sagittarius A East it is much larger  ($\sim 140~\rm km^{-2}~yr^{-1}$).
As far as the Crab nebula is concerning, the event rates turn out to be close
to the ones expected from the atmospheric neutrino background.
With these predicted fluxes, RX J1713.7-3946 and 
Sagittarius A East are sources of interest also for telescopes 
of smaller dimensions than 1 km$^2$, such as the ANTARES telescope 
(Amram et al.~1999) and AMANDA-II (Ahrens et al.~2004a). 
\section{Pulsars in high density regions}
\label{sec:pulsars}
The lifetime of massive stars which explode as supernovae, with subsequent
formation of pulsars, is relatively short, of the order of $10^6-10^7$ yrs. 
Therefore,
pulsars are likely to appear inside or close to massive stellar
associations where the amount of distributed matter is very high. Particles 
accelerated by pulsars can be trapped by this high density medium, which is
characterized by magnetic fields of the order of 
$10^{-5}-10^{-4}$ G, or even $\sim 10^{-3}$ G inside huge molecular
clouds (Crutcher~1999). The largest stellar associations in our Galaxy contain
hundreds early type massive (OB) and Wolf-Rayet (WR) stars. 
Hence, the average formation rate of pulsars 
inside the massive associations can be estimated to be about $10^{-4}$ 
per year. Therefore, it is likely that on average a
relatively young pulsar, with an age of $\sim 10^4$ yrs, should   
be present in every massive association. Such 
pulsars should be efficient accelerators of particles up to very high energies,
as it is proved by their pulsed $\gamma$-ray emission and surrounding 
non-thermal nebulae. 
\subsection{The Galactic Centre dense region}
\label{sec:galcen}
The most interesting site for the occurrence of the processes
described above is the huge concentration of mass in the 
inner $\sim 50$ pc
around the Galactic Centre (GC), with total mass of $\sim 10^6$ M$_\odot$, 
average density $10^2$ cm$^{-3}$, and magnetic field $3\times 10^{-5}$ G. 
This region contains
also a huge molecular cloud with radius 10 pc, density $10^3$ cm$^{-3}$, 
and magnetic field $10^{-4}$ G. The GC is also a site where high 
energy processes occur. For example, the EGRET detector on board of the
{\it Compton Gamma-Ray Observatory} 
satellite has detected an intense $\gamma$-ray source with luminosity 
$2\times 10^{37}$ erg s$^{-1}$ (Mayer-Hasselwander et al.~1998).
There are also some indications of TeV $\gamma$-ray emissions from this region
(Kosack et al.~2003, Tsuchiya et al.~2003). 
The AGASA collaboration reported the existence of an extended excess of
cosmic rays (CR) of energy around $\sim 10^{18}$ eV from the direction
close to the GC 
(Hayashida et al.~1999). This excess has been confirmed by the analysis 
of the SUGAR data (Bellido et al.~2001). 

Recently, Bednarek (2002) has considered a  
scenario in which energetic pulsars accelerate nuclei inside (or close to) this
high density region to energies observed by the above mentioned cosmic ray
experiments. Such a pulsar should have an initial period 
$P_0 = 10^{-3}P_{0,ms}$ s and surface magnetic field $B = 10^{12}B_{12}$ G, 
which fulfill the condition $B_{12}P_{0,ms} > 0.4$
in order to accelerate hadrons above 
$10^{18}$ eV. As an example, a pulsar 
with $P_0 = 2$ ms and $B = 6\times 10^{12}$ G has been chosen. 
The nuclei accelerated by the pulsar, as discussed in the model by 
Bednarek \& Bartosik (2003), escape from the pulsar nebula and are partially 
captured by the high density region in the GC. They partially disintegrate
in collisions with matter, thus injecting high energy neutrons which are 
responsible for the observed excess of cosmic rays with energy 
$\sim 10^{18}$ eV. Moreover,
these nuclei produce $\gamma$-rays and neutrinos. Bednarek (2002) has 
also investigated the likely parameter range for the GC medium, 
considering the case of hadrons trapped inside the dense molecular cloud
with density $10^3$ cm$^{-3}$, radius $10$ pc, and magnetic field
$10^{-4}$ G, and a larger region around the GC with density $10^2$ cm$^{-3}$,
radius 50 pc, and magnetic field $3\times 10^{-5}$ G.
From the normalization of the
calculated neutron flux to the observed flux of CR particles in excess, he
predicts the fluxes of TeV $\gamma$-rays, which should be detected by the 
next generation of Cherenkov experiments, and the fluxes of neutrinos at the 
Earth. The neutrino spectra are characterized by power laws with a
spectral index of $\sim -2.5$ from $\sim 10-100$ TeV up to $\sim 10$ PeV.
It is found, considering the probability of detecting muons produced by 
such neutrinos (Gaisser \& Grillo 1987) and the neutrino 
absorption by the Earth (Gandhi~2000), that a few up to several
muon neutrino events should be observed in a 1 km$^2$ neutrino detector per year
for the horizontal (H) and the vertical (V) directions from a pulsar which was formed 
$(3-10)\times 10^3$ yrs ago (see Tab.~\ref{tab2} for the details).  
\begin{table} 
      \caption{$\nu_\mu$ event rate per year in a km$^2$ detector
 from the Galactic Centre for an $E^{-2.6}$ spectrum between 10-100 TeV and
$\sim 10$ PeV} 
\begin{center}
	 \begin{tabular}{|c|c|c|c|} 
	    \hline 
	     medium $\Rightarrow$    & GC extended & dense molec. \\ 
	     pulsar age $\Downarrow$ & ~~~~region & ~~cloud       \\
	     \hline 
	     $3\times 10^3$ yrs (H - V)  & 3.8 - 2.0 &  16 - 11   \\   	    
            \hline 
	     $10^4$ yrs (H - V)          & 8.8 - 5.3 &  30 - 23   \\
 	    \hline
	 \end{tabular}
\end{center} 
	 \label{tab2} 
     \end{table} 

The observed  CR excess at $\sim 10^{18}$ eV is also interpreted as due to
the last (or last few) Gamma Ray Bursts (GRBs) which occurred in our Galaxy 
(Biermann et al.~2004). If the
rate of GRBs is of the order of one per $10^6$ yrs, then hadrons accelerated by the GRB
relativistic shock convert partially to neutrons which decay into protons
inside the Galaxy, and
after some time protons are partially converted into neutrons in collisions with matter. The relative flux of neutrinos to the flux of neutrons, produced in these hadronic collisions, with energies below $\sim 10^{18}$ eV has been 
estimated. 
It should produce similar event rates in a 1 km$^2$ neutrino detector as those estimated by 
Bednarek~(2002).

The relativistic hadrons can also be accelerated by the shock waves of the supernovae
which appeared in the central high density region. One of such shell-type SNR
associated with Sgr A East, with an age of $\sim 10^4$ yrs, has been recently 
considered as responsible for the unidentified EGRET source at the GC 
(see Fatuzzo \& Melia~2003).
These relativistic hadrons should interact with the high density medium ($10^3$ 
cm$^{-3}$) producing $\gamma$-rays and neutrinos. However, expected neutrino fluxes 
at TeV energies have not been estimated yet in the context of this model. 
\subsection{The massive stellar associations}
\label{sec:massive}

The largest massive stellar association, at a relatively small distance of 1.7 kpc,
is Cygnus OB2. Its total mass derived from the CO survey is $\sim 3.3\times
10^5$ M$_\odot$ (Butt et al.~2003) and the total stellar mass is $(4-10)\times 10^4$
M$_\odot$ (Kn\"odlsedler~2000). Moreover, it contains many massive stars of the O type,
$\sim (120\pm 20)$ (Kn\"odlsedler~2000).
Recently, Cygnus OB2 became also an interesting high energy source due to the reports of
detection of a few $\gamma$-ray EGRET sources (e.g. Mori et al.~1997, 
Lamb \& Macomb~1997), a TeV $\gamma$-ray source by HEGRA
(Aharonian et al.~2002), and an excess of CR particles at energies $\sim
10^{18}$ eV by AGASA
(Hayashida et al.~1999). Therefore, it is also an interesting 
potential source of high energy neutrinos.

Torres et al. (2004) propose that TeV $\gamma$-rays and neutrinos 
can be produced by CRs through interactions with the nuclei in the
innermost parts of the winds of massive O and B stars.
This model would explain the TeV emission observed by HEGRA
in the vicinity of Cygnus OB2 (Aharonian et al.~2002).
The main acceleration region of CRs would be the
outer boundary of the supper-bubble produced by the core of
the association of OB stars.
Moreover, only CRs with energies $>1$ TeV are able to penetrate dense 
inner regions of the winds from outside since for them the diffusion 
timescale in the wind is shorter than the convection time scale for 
likely parameters of the massive stars. 
This fact explains why a GeV $\gamma$-ray source related to the reported 
TeV $\gamma$-ray source by HEGRA has not yet been observed. 
On the other hand, a nearby EGRET source, 3EG J2033+4118, is coincident with
the center of the association, where the GeV emission might be produced
in the terminal shocks of powerful stars existing therein, in the
colliding wind binary system Cyg OB2 No. 5 or in a combination of these
scenarios.
In order to explain the observed flux from the unidentified 
TeV $\gamma$-ray source, the model requires an enhancement of 
the CR flux in the region of Cygnus OB2 of the order of 300 compared to
the CR flux in the vicinity of the Earth,
if the ISM density is  of the order of 0.1 cm$^{-3}$.
The predicted neutrino signal from Cyg OB2 above 1 TeV in the IceCube
neutrino detector should be a factor $\sim 1.8$ larger than the 
atmospheric neutrino background in search bins of $1^{\circ} \times 1^{\circ}$.

Another model able to explain possible high energy phenomena in Cygnus OB2 
assumes the 
birth of a very energetic pulsar about $10^4$ yrs ago inside this
association (Bednarek~2003b). Taking into account the number of massive stars
inside Cygnus OB2 ($\sim 100$) and their characteristic lifetime (of a few $10^6$ yrs), 
such pulsars are likely to be produced with the required rate of one per 
$\sim 10^4$ yrs. 
This pulsar, with reasonable parameters, such as an initial period of 2 ms
and a surface magnetic field of
$6\times 10^{12}$ G, would be able to accelerate heavy nuclei to Lorentz factors 
of $\sim 10^{9}$ soon after formation. Part of these nuclei can be captured for a long
time in the dense regions of Cygnus OB2, characterized by a strong magnetic
field ($\sim 10^{-4}$ G), producing relativistic neutrons in collisions with matter. 
These neutrons can be responsible for the observed CR excess observed in 
AGASA (Hayashida et al.~1999). On the other hand, the 
pulsed emission of this Vela type pulsar, 
$\sim 10^4$ yrs old, could be 
responsible for the observed GeV $\gamma$-ray source.
In fact, the level of $\gamma$-ray emission and the shape of the spectrum 
of the Vela and PSR 1706-44 pulsars are consistent 
with the observations of the EGRET source 3EG J2033+4118 in the direction
of Cyg OB2. 
The TeV $\gamma$-ray emission from the nebula
surrounding this pulsar can in turn explain the HEGRA TeV $\gamma$-ray source  
(see for detailed modeling Bednarek \& Bartosik~2003).
However, the model considered by Bednarek~(2003b) predicts that the fluxes of muon 
neutrinos, produced by nuclei within the pulsar nebula which in turn are captured by the 
massive association, are too low to be detected by a 1 km$^2$ neutrino detector ($\sim 0.5$ 
neutrino events from the pulsar nebula and $\sim 0.14$ from the dense
region of the Cygnus OB2 per year).

It has also been suggested that relativistic nuclei may suffer strong 
photo-disintegrations in the radiation field present inside the massive association 
Cyg OB2 (Anchordoqui et al.~2003a). This soft photons are produced by the above
mentioned 
population of the massive stars and the warm molecular clouds. 
Neutrons extracted from the nuclei,
with energies below $\sim 2\times 10^{17}$ eV,  decay into antineutrinos
during their propagation towards the Earth along distances smaller than
1.7 kpc. A few antineutrinos of all flavors (since the electron 
antineutrinos from neutron decay are expected to oscillate into other
flavors) might be detectable per year by a 
1 km$^2$ detector if the spectrum of nuclei inside the Cyg OB2 is very
steep $\propto E^{-3.1}$ in the PeV-EeV region
(Anchordoqui et al.~2003a). 
Nevertheless these calculations do not consider that
the interactions of nuclei with matter 
(as considered by Bednarek~2003b) might be more important
than those with radiation. 
If the total amount of gas inside the core of Cyg OB2 (within 10 pc) 
is $\sim 3\times 10^4$ M$_\odot$ (corresponding to an average density of 
$\sim 300$ cm$^{-3}$), then  
neutrinos produced in collisions of nuclei 
with matter would dominate over 
neutrinos from decays of neutrons, produced in photodisintegrations of nuclei. 
This would be due to the larger (by about an order of magnitude) 
cross sections for collisions of nuclei with
matter, with respect to radiation. 

\section{Neutron stars in binary systems}
\label{sec:binary}
The calculations of neutrino fluxes from binary systems have been reviewed
by Gaisser et al. (1995). 
Most of the earlier models assumed that particles are accelerated due to 
the energy generated during the accretion of matter onto a slowly rotating
neutron star (e.g. Gaisser \& Stanev 1985, Kolb et al.~1985, Berezinsky et al.~1985). 
Those models predict high neutrino fluxes since they are normalized to the
measurements of VHE $\gamma$-ray emissions from Cyg X-3, 
still nowadays not well understood.
In another model, Harding \& Gaisser (1990) consider hadrons accelerated 
in the pulsar wind shock taking as a reference the massive binary Cyg X-3.
This model predicts detectable fluxes of neutrinos in a 0.1 km$^2$ detector
(Gaisser et al.~1995) only for the case of eclipsing binaries, when 
the massive star is between the source of relativistic hadrons and the observer.

A similar scenario has been more recently considered by Bartosik et al. 
(2003), i.e. a binary system
containing an energetic pulsar and a massive stellar companion, 
with the purpose  
to calculate the neutrino fluxes produced by pulsar accelerated nuclei. 
Provided that the binary system is compact enough and the star is luminous,
the nuclei suffer photodisintegrations in the thermal radiation field of 
the massive star. The neutrons from disintegration move ballistically, some 
of them interact with the massive star matter. On the contrary, the
nuclei which survived disintegration and the secondary protons are injected 
through the shock structure into the magnetic field in the massive star wind 
region. Bartosik et al. calculate the paths of charged hadrons 
in the massive star magnetic field.
It is found that some of these charged hadrons can fall at large angles 
(larger than expected from the simple linear propagation) onto the 
massive star, thus
producing neutrinos. The neutrino fluxes and spectra produced by neutrons 
and charged hadrons are calculated for different locations of the observer 
with respect to the orbital plane of the binary system. It is found that 
interesting fluxes of neutrinos are emitted at angles larger than those
intercepted by the massive star. Therefore, not only eclipsing binaries 
might become detectable sources of neutrinos by the future telescopes.  

Slowly rotating, accreting neutron stars have been also considered recently 
as a source of neutrinos. It is proposed that if a neutron star is strongly 
magnetized and spins down slower than the accretion disk, then an accelerating
gap should appear in the pulsar magnetosphere above the accretion disk 
(Cheng \& Ruderman~1989). Hadrons can be accelerated in such a gap up to $\sim 100$ TeV,
if the disk luminosity is of the order of $10^{37}$ erg s$^{-1}$, 
and the surface magnetic 
field of the neutron star close to $10^{12}$ G. 
In the frame of this model, Anchordoqui et al. (2003b) estimate the 
neutrino flux produced in cascades initiated by relativistic hadrons in 
the matter of the optically thick accretion disk. The authors predict a 
neutrino signal above the atmospheric neutrino background in a
region of $1^{\circ} \times 1^{\circ}$ around the source 
(signal to noise ratio $\approx 1.7$) between 300 GeV and 1 TeV from the binary system A0535+26. 
The predicted neutrino flux, in units of 
$\rm GeV^{-1}~ m^{-2}~ s^{-1}$, is given by 
\begin{eqnarray}
  \frac{dN_\nu}{dE_\nu} & = &  4.2 \pm 0.3 \times 10^{-3}\, \rm 
\left(\frac{E_\nu}{1 \mbox{GeV}}\right)^{-2.35 \pm 0.02}~~
   (E_\nu < 10^3~ GeV) \nonumber \\
& &  \label{acc}
\end{eqnarray}    
\noindent
The signal should appear when the neutron star is close to the periastron 
due to 
the formation of a dense accretion disk. Therefore, it should
be modulated with the period of the binary system. The neutrino signal
from A0535+26 should be above the sensitivity of the IceCube detector
during a few binary system reversals.
\section{Microquasars}
\label{sec:mq}

Microquasars are galactic X-ray binary systems, which exhibit 
relativistic radio jets. Therefore, similar processes of neutrino production
by relativistic particles as occurring in active
galactic nuclei (AGNs) can be considered, e.g. proton initiated cascade model
in the jet (Mannheim \& Biermann 1992, M\"ucke \& Protheroe~2001), in the 
radiation of 
an accretion disk (Bednarek \& Protheroe 1999), or interaction of 
neutrons with 
the matter of an accretion disk (Nellen et al.~1993). 

Recently, Levinson \& Waxman (2001) have proposed a model for neutrino 
production in microquasars, similar to the AGN 
proton initiated cascade model of Mannheim et a.~(1992). 
The authors argue that hadrons can be accelerated up to $\sim 10^{16}$
eV in the inner part of the jet by internal shocks in the jet.
Pions are produced in collisions of hadrons with the external X-ray photons,
and with the synchrotron photons from the jet produced by leptons 
accelerated in this same 
shock. The neutrino fluxes, expected in the energy range 1-100 TeV, are
very high, especially from persistent sources, e.g.,
$\sim 10^3$ neutrino event rates in a 1 km$^2$ neutrino detector in the case 
of SS433. For the case of bursting sources, such as GRS 1915+105, the
expected sensitivity of a 1 km$^2$ detector would require observation of 
about 30 outbursts for detection. 
In a subsequent paper, Distefano et al. (2002) estimated the expected neutrino 
event rates in a 1 km$^2$ detector from the large population of known 
microquasars,
assuming that relativistic protons take 10$\%$ of the jet power. 
The authors conclude that persistent microquasars, such as SS433 or 
GX 339-4, should produce more than 100 neutrino events in such a 1 km$^2$ detector.
Therefore, some of the neutrino events could be observed by 
AMANDA-II. Also neutrinos produced in a long duration flare from
microquasars similar to, e.g., XTE J1118+480, Cyg X-3 or GRO J1655-40,
should produce a few neutrino events in a 1 km$^2$ detector. 
\noindent
In the following, we report the neutrino fluxes given by Distefano et al. 
(2002, private communication):
\begin{eqnarray}
\frac{dN_\nu}{dE_\nu} & = & \frac{K}{(4+5\ln20)}~F_\nu~
\left(\frac{E_\nu}{1 \mbox{GeV}}\right)^{-1}~(\rm E_\nu < 5~10^3~GeV) \nonumber \\
 & = & \frac{5000~K}{(4+5\ln20)}~F_\nu~
\left(\frac{E_\nu}{1 \mbox{GeV}}\right)^{-2}~(\rm E_\nu > 5~10^3~GeV)
\label{mqflux}
\end{eqnarray}
being $\rm K=10^4/1.6$ a constant needed to adapt Distefano et al. 
flux units (in erg cm$^{-2}$ s$^{-1}$) 
to (GeV$^{-1}$ m$^{-2}$ s$^{-1}$). The normalization factor
$(4+5\ln20)$ is due to the presence of a knee in the spectrum. 
$F_\nu$ is a number characterizing the specified source, and is 
calculated in the model. Some selected sources are listed in Tab.~\ref{tab3},
along with the expected $\nu$ fluxes.

\begin{table}
\caption{List of microquasars considered in Distefano et al. (2002), 
including expected neutrino fluxes in units of $\rm erg~s^{-1}~cm^{-2}$.}
\begin{center}
\begin{tabular}{|rl|}
\hline
Source &  $\rm F_\nu$ \\
\hline\hline
XTEJ1748-288 &  $3.07\times10^{-10}$\\
CygnusX-3    &  $4.02\times10^{-9}$\\
LS5039 &   $1.69\times10^{-12}$\\
GROJ1655-40 &        $7.37\times10^{-10}$\\
GRS1915+105 &        $2.10\times10^{-10}$\\
CircinusX-1  &        $1.22\times10^{-10}$\\
XTEJ1550-564 &        $2.00\times10^{-11}$\\
V4641Sgr(1)  &       $2.25\times10^{-10}$\\
V4641Sgr(2) &         $3.25\times10^{-8}$\\
ScorpiusX-1 &         $6.48\times10^{-12}$\\
SS433  &              $1.72\times10^{-9}$\\
GS1354-64 &           $1.88\times10^{-11}$\\
GX339-4    &         $1.26\times10^{-9}$\\
CygnusX-1  &         $1.88\times10^{-11}$\\
GROJ0422+32 &        $2.51\times10^{-10}$\\
\hline
\end{tabular}
\end{center}
\label{tab3}
\end{table}

In the case of microquasars containing close massive stars of OB type, the 
relativistic hadrons accelerated in the jet can also interact with 
the dense wind of the massive star.
In their model, Romero et al. (2003) consider as likely parameters of such 
a binary system the following ones:
mass loss rate of massive star $\sim 10^{-5}$ M$_\odot$ yr$^{-1}$,
wind velocity
$2500$ km s$^{-1}$, radius of the massive star 35 $R_\odot$,
mass of the black hole 10 $M_\odot$, accretion rate onto the black hole $10^{-8}$ M$_\odot$ yr$^{-1}$. 
They conclude that, if the spectrum of the accelerated protons is of 
power law type, with spectral index close to 2, and it extends up to 100 TeV, 
the fluxes of neutrinos 
produced in collisions of jet accelerated hadrons with the matter of the wind 
might be a factor of 3 above the atmospheric neutrino background 
in the energy range 1-10 TeV,
if the efficiency of energy conversion of the accretion disk to the 
relativistic protons is equal to $10^{-2}$.

Simultaneous observations of 
the X-ray and neutrino emission from specific microquasars 
can help to distinguish between these 
two models or at least estimate their relative importance.

\section{Magnetars}
\label{sec:magnetars}
Magnetars are isolated neutron stars with surface dipole magnetic fields 
much higher than in ordinary pulsar, typically $\sim$ 10$^{15}$ G. Such objects
manifest themselves in the form of soft gamma-ray repeaters (SGRs) and
anomalous X-ray pulsar (AXPs). 
The prime distinction between a pulsar and such 
a magnetar is that the X-ray and particle emissions from the magnetar
are powered not by rotation, as in common pulsars, but also 
by the decaying magnetic field. This involves both internal heating and
seismic activity that shakes the magnetosphere and accelerate particles.
This gradual release of energy is punctuated by intense outbursts that are 
most plausibly triggered by a sudden fracture of the neutron star's 
rigid crust.

In a recent paper, Zhang et al. (2003) propose that young fast-rotating 
magnetars are likely to be significant TeV neutrino sources, and therefore
interesting targets for the planned km$^3$ neutrino detector.
In their model, Zhang et al. (2003) argue that the maximum potential drop
through the magnetar magnetosphere might be high enough to 
accelerate protons to energies above the photomeson threshold.
These protons will then interact 
with the thermal radiation from the neutron star 
surface, thus producing pions and neutrinos. 
In order to accelerate protons to the required energies,
the parameters of the magnetar have to fulfill the following condition

\begin{equation}
P < (2.4 - 6.8~{\rm s})~B_{p,15}^{1/2} R_6^{3/2}
(1-\rm cos\theta_{p\gamma})^
{1/2}
\label{period}
\end{equation}
\noindent
where $P, B_p$, and $R$ are the rotation period, surface
(dipolar) magnetic field, and stellar radius, and $\theta_{p\gamma}$ is
the interaction angle between the photon and the proton.
The range of periods in the right-hand side of eq.(\ref{period})
defines a ``neutrino death valley'' in the magnetar $P-\dot P$
(or $P - B_p$) space. Photomeson interactions and neutrino emission cease
when the magnetar crosses this valley from left to right during its
evolution (see Fig.1 of Zhang et al. paper). 
Four magnetars are found to be close to
the valley, which means that under favorable conditions, they are high
energy neutrino emitters.  For an on-beam observer, and assuming that 
the neutrino luminosity is beamed into a solid angle $\Delta \Omega_\nu$
around the polar axis, Zhang et al. find that the neutrino 
flux at Earth is
\begin{eqnarray}
\phi_{\nu} & \simeq &  2.1 \times 10^{-12} {\rm cm^{-2} \, s^{-1}} 
\left( \frac{\Delta \Omega_\nu}{\rm 0.1}\right)^{-1} 
\left( \frac{\eta_p}{\rm 0.5}\right)^{2} 
\left( \frac{f_{cool}}{\rm 0.25}\right) \nonumber \\
& & B_{p,15}^3 R_6^{10} \left( \frac{P}{\rm 5 s}\right)^{-6} 
\left( \frac{kT}{\rm 0.5 keV}\right)^{4}\left( \frac{D}{\rm 5 kpc}\right)^{-2} 
\left( \frac{\epsilon_\nu}{\rm 2 TeV}\right)^{-1} \nonumber \\ 
\label{magn_new}
\end{eqnarray} 
\noindent
where $D$ is the distance to the source. Moreover $f_{cool}$ is the pion 
cooling factor, $kT \sim$ (0.4-0.6) keV is the observed blackbody temperature 
for SGR/AXP quiescent emission, and $\eta_p$ parametrizes the uncertainty
in the utilization of the polar cap unipolar potential. 
Zhang et al. have found that for on-beam detections, SGR1900+14 and 
1E1048-5937 have substantial neutrino fluxes, making them interesting 
targets for the planned large area Cherenkov detectors.
Unfortunately, event rates strongly depend on this beaming angle. In fact,
a smaller  $\Delta\Omega_\nu$ increases event rates, as neutrinos flux
depends on (1/$\Delta\Omega_\nu$), but decreases the probability of 
on-beam detection.
Moreover, we notice that the neutrino flux spectral index is 
the same for all sources, while the normalization factor changes as it depends
on the distance, the rotation period and the surface magnetic field of the 
source.

The estimated neutrino event rate from one of the more likely 
neutrino source, SGR1900+14 should produce 
$\sim (1.5-13)(0.1/\Delta\Omega_\nu)$ events in 1 yr and in a km$^2$ detector,
if the protons are
accelerated with the efficiency corresponding to the Goldreich \& Julian 
(1969) density. Such a rate is above the sensitivity of a 1 km$^2$ 
detector, provided that the observer is located within the cone of neutrino 
emission. 
There might be some problems with acceleration of hadrons to 
the required energies in this model.
The authors estimate the mean free paths for pion and $e^\pm$ pair
production in hadron-photon collisions for typical parameters of thermal 
radiation field expected in magnetars and conclude that the electric field in 
the acceleration region is not saturated by created pair plasma. However,
it is assumed that hadrons are injected with the Goldreich and Julian (GJ, 
1969) density and the number of $e^\pm$ pairs produced by hadrons per one 
pion should be of the order of ten (Zhang et al.~2003), i.e. ten times 
above GJ density. 
These pairs are created in strong radiation and electric fields so they 
should initiate very efficient cascades since the mean free paths for the 
ICS and subsequent $e^\pm$ pair production are approximately one thousand
times shorter than for pion production. Therefore the number of secondary 
pairs in such dense radiation and strong electric fields can be enormous 
(see e.g. Bednarek \& Kirk 1995).

Another problem of that model is that, in the above discussions about magnetar
gap acceleration, a dipole configuration is assumed, whilst a magnetar 
magnetosphere is certainly 
not dipolar. On this point, 
Thompson et al. (2002) argue that the SGR/AXP 
phenomenology can be consistent with the hypothesis that the magnetar 
magnetosphere is instead globally twisted. 
\section{Summary and conclusions}
\label{sec:conc}
In general we can distinguish transient, persistent, and periodic neutrino sources.
Some of these sources should produce 
strongly collimated neutrino beams, e.g. supernova models 
(Protheroe et al.~1998), binary system models (Anchordoqui et al.~2003b, 
Bartosik et al.~2003), or magnetar model (Zhang et al.~2003). Therefore, 
a specific 
geometry is required in order to have a chance to detect them. 

The neutrino fluxes expected from transient sources such as the supernovae
can produce from a hundred up to a few thousands of neutrino events in a 1 km$^2$ 
detector during the first year after supernova explosion within our Galaxy,
provided that the accelerated hadrons take a significant part of the
available supernova energy
(Waxman \& Loeb~2001, Beall \& Bednarek~2002, Nagataki~2004).  However, 
a more realistic estimation of the hadron acceleration efficiency, 
which is based on the condition that the observed cosmic 
ray flux in the Galaxy should not be exceeded (Blasi et al.~2000, Beall \&
Bednarek~2002), produce
significantly reduced neutrino event rates in a 1 km$^2$ detector. A few
events can be expected in the case of pulsar formation
with typical values of the surface magnetic fields of radio pulsars, but
with initial periods of the order of a few milliseconds. If a pulsar with 
extremely high surface magnetic field is born in a supernova explosion, 
a so called magnetar, then the estimated event rates might be higher. However, 
in this case the pulsar can additionally accelerate the supernova envelope, 
dropping 
faster the column density of matter for relativistic hadrons and consequently 
lowering the expected neutrino event rates. This effect has not been taken into account 
in the calculations presented by Beall \& Bednarek~(2002) and Nagataki~(2004). 
In general, detection of 
neutrinos from the early stage of supernovae in a reasonable time in not very promising 
due to the low rate of supernova explosion in the Galaxy $\sim 10^{-2}$ yr$^{-1}$. 
Estimated event rates from Galactic supernovae do not give also much hope of 
detecting such supernovae in the nearby galaxies. For example,
the expected event rates from the supernova SN1987A in LMC, at distance of $\sim
5$ times larger than a typical Galactic supernova, are below or at the level
of sensitivity of a 1 km$^2$ detector in most of the considered 
realistic models. 

The microquasar and magnetar models (Levinson \& Waxman~2001,
Zhang et al.~2003) predict efficient production of neutrinos during the outbursts 
observed in the radio and/or X-ray bands. The predicted strong 
correlation between the expected neutrino fluxes and this low energy emission 
should eventually help in extracting the neutrino signal from the atmospheric 
neutrino background. This will require 
long multi-wavelength campaigns since the expected detection of the neutrino 
signal 
from , e.g. GRS 1915+105, will need the observation of about 30 outbursts. 

More promising detection of neutrinos is expected from persistent sources such as the 
PWNe and the shell-type SNRs, provided that a significant part of their  
TeV $\gamma$-ray emission have hadronic origin. The expected neutrino event rates from 
these sources are relatively low in a 1 km$^2$ detector, of the order of a few events 
yr$^{-1}$ (Alvarez-Muniz \& Halzen~2002, Bednarek~2003a, Amato et al. 2003), but we know exactly where to 
look for them. 
Therefore, even a relatively weak signal from the direction of the
Crab Nebula or TeV $\gamma$-ray shell-type SNRs will give important constraints
on hadronic processes.   
Another likely persistent neutrino source is expected in the Galactic Centre region. 
If excess of CRs from the Galactic Centre at $\sim 10^{18}$ eV, found in the AGASA and SUGAR 
data, is confirmed (Hayashida et al.~1999, Bellido et al.~2001), then it should be accompanied 
by a neutrino source able to produce up to several neutrino events in a 1 km$^2$ 
detector per year (Bednarek~2002, Anchordoqui et al.~2003a, Biermann et al.~2004). 
In this case, neutrino fluxes would be correlated to
the measured CR flux.

Also the models involving hadron acceleration inside massive binary systems
are expected to be persistent and periodic (with the orbital period)
neutrino sources. Bartosik et al. (2003)
consider in more detail the old scenario in which energetic pulsar is responsible for
acceleration of hadrons which interact with the massive companion matter. 
The model predicts that  detectable neutrino
fluxes can be observed, also from non-eclipsing binary systems,
due to the propagation of relativistic hadrons in the 
magnetic field of the massive star, which allows them to fall onto the star 
at large
angles to the plane of the binary system. Anchordoqui et al.~(2003b) argue that
the relativistic hadrons impinging on the surface of the accretion disk around
the neutron star, which is companion of the massive star, can produce
a neutrino signal above the atmospheric neutrino background  
in a 1 km$^2$ detector near the periastron for very dense 
accretion disks, e.g. in the case of A0535+26.

Models for astrophysical neutrino sources in our Galaxy usually, very optimistically, 
assume efficient transfer of the energy generated inside the source
into relativistic hadrons. It looks that the acceleration 
efficiency is going to be constrained for some sources by the upper 
limits derived on the basis of AMANDA observations (Ahrens et al.~2003, Hauschildt et al.~2003). For example, 
the microquasar model (Distefano et al.~2002) prediction
is about a factor of 2 below the AMANDA-II 90\% c.l. upper limit
for the persistent source SS433 (Ahrens et al.~2004a).

In general, models which postulate the
production of neutrinos in hadron collisions with the radiation field already
inside the acceleration region 
(e.g. the early stage of the model by Beall \& Bednarek~(2002), the microquasar
model by Levinson \& Waxman (2001), or the magnetar model by Zhang et al.~(2003)) 
may have problems with efficient
conversion of energy from the acceleration mechanism into relativistic hadrons.
This is due to the fact that the threshold for $e^\pm$ pair production is about
two orders of magnitude below the threshold for pion production and the cross section
for $e^\pm$ production is significantly larger than that for pion production
in the case of hadron-photon collisions. Therefore, pair
production is more efficient than pion production by protons, 
with the possibility to have $e^\pm$ pair domination of accelerated plasma. Such 
$e^\pm$ pair domination of the plasma during the acceleration process in the electric 
field (induced in reconnection regions) has been 
discussed by Bednarek \& Kirk~(1995) and in the case of shock accelerated particles by
Mastichiadis \& Kirk (1995). As a result, most of the energy from the acceleration 
mechanism is transferred to leptons and it is radiated in the form of electromagnetic 
radiation but not of neutrinos.
In the case of models which postulate production of neutrinos in hadron collisions  
with matter the situation is more promising. The cross section for
$e^\pm$ pair production in collisions of two hadrons is significantly lower (even for
nuclei as heavy as iron) than the cross section for pion production.
Therefore, the domination in the acceleration region by the $e^\pm$ plasma is not 
possible, provided that other processes, e.g. leptonic, do not contribute significantly. 
Moreover, in these models the region of hadron acceleration is usually different than 
the region of their interaction with matter.

\section*{Acknowledgments}
WB would like to thank for the support and hospitality during his visit in INFN 
Sezione do Catania and for the support by the Polish KBN grant No. 5P03D 025 21.
TM would like to thank F. Vissani for useful discussions.

\newpage
.
\begin{sidewaystable}
\renewcommand\tabcolsep{.05pc} 
\renewcommand{\arraystretch}{.5} 
\caption{Brief summary of the models described in this review paper.
The source type and distance are indicated, the process that produces 
neutrinos, the duration of the neutrino emission (D) or/and the source age
(A), the relevant energy region for producing the given event rates in a km$^2$
detector per year and the neutrino flux spectrum.}
\begin{tabular}{|c|c|c|c|c|c|c|c|} 
\hline \hline 
Source type       & distance &  process & duration of & 
$\rm E_\nu$ & $dN/dE_{\nu}$& $\rm N_{\nu_\mu}$ & Ref.\\
                  &  (kpc)          & & emission (D) or & (GeV) & (GeV$^{-1}$
m$^{-2}$ s$^{-1}$)& ($km^{-2} yr^{-1}$)  & \\
                  &          &          & source age (A) & & & & \\
\hline 
{\bf supernovae} & 10 &                         &  & & & &\\
shocks           &    & $pp$                    & $\sim 1$ hr (D) $\sim 10$ hrs (A)& $\lesssim \rm 10^3$ & $\propto$ E$^{-2}$ & $\sim \rm 100$ &  Waxman \& Loeb (2001) \\
pulsars          &    & $Fe-\gamma\rightarrow n$ & $\sim 0.1$ yrs (D)&
$\sim \rm (10^2-10^6)$ & $(10^{-5}-10^{-4})$E$^{-(1.7 \div 1.8)}$($<10^4$GeV) & $50-10^3$ & Protheroe et al. \\
                 &    &  $np$                    &                &
& $(10^{3}-10^{6})$ E$^{-(3.6 \div 3.7)}$($>10^4$GeV) &  & (1998) \\
                 &    &                          &                &
& (for P = 5-10 ms) & & \\
                 &    & $N-\gamma$+Fe-N & $10^{4}-10^{7}$ s (D)& $\sim \rm 10^5-10^8$ & $\propto$ E$^{-1}$ &  $\sim 10-10^3$ & Beall \& Bednarek (2002) \\
                 &   &$pp$ & $\lesssim 100$ yrs (D)& $\sim 10-10^8$ & & $\lesssim 10^3$ & Nagataki (2004) \\
\hline
{\bf Plerions}   & 0.5-4.4 & $pp$ & $\lesssim 3\cdot 10^4$ yrs (A)& $< 10^3-10^5$& $(0.15-1.85) \cdot 10^{-10}$E$^{-(2.4 \div 2.6)}$ &$\sim 1-12$ & Guetta \& Amato (2003,\\
                 &    &                          &                &
&  &  & private communication) \\
                 &         & $Fe-p$ &                             & $\sim \rm 10^{3}-5 \cdot 10^5$ & & $\lesssim 1$  & Bednarek (2003a) \\
Crab             & 2       & $Fe-\gamma\rightarrow n$ & $\sim 10^3$ yrs (A)&$\sim \rm 10^{3}-5 \cdot 10^5$ & $1.7 \cdot 10^{-6}$ E$^{-1.6}$($<10^4$GeV) & few  & Bednarek \& Protheroe \\
                 &         & $np$       &                    &  & $1.5 \cdot 10^{2}$ E$^{-3.6}(>10^4$GeV)& & (1997) \\
                 &         & $Fe-p$ &   &$\sim \rm 10^{3}-5 \cdot 10^5$ & $4 \cdot 10^{-8}$ E$^{-1.44}$($<6\cdot 10^4$GeV) & $\sim 1$  & Bednarek (2003a) \\
                 &         &        &   &                               & $3.1 \cdot 10^{14}$ E$^{-6.11}$($>6\cdot 10^4$GeV) & $\sim 1$  & Bednarek (2003a) \\
                 &         & $pp$   &                    & $10-10^6$ &$\propto$ E$^{-2.25}$  & $\sim 4-14$ & Amato et al. (2003) \\
\hline 
{\bf shell SNRs}    &    &                         &   & & & &\\
SNR RX J1713.7-3946 & 6 & $pp$& $\sim 10^3$ yrs (A)& $\lesssim 10^4$ & $4.1\cdot 10^{-4}$ E$^{-2}$&$\sim 40$ & Alvarez-Muniz \& Halzen \\
Sgr A East          & 8 & & $\sim 10^{4}$ yrs (A) &$\lesssim 10^5$ & & $\sim 140$ & (2002) \\
\hline
{\bf Pulsars + clouds} & & & & & & & \\
Galactic Centre        & 8 &$Fe-p$ & $\sim 10^{4}$ yrs (A)& $10^4-10^7$ & $0.166\cdot$ E$^{-2.64}$ & $\sim 2-30$ & Bednarek (2002) \\
Cygnus OB2             & 1.7 & $pp$& $\sim 10^{6}$ yrs (A)&$\gtrsim 10^3$& $\propto$ E$^{-2}$ &few &Torres et al. 2004\\
                       &     & $Fe-p$ & $\sim 10^{4}$ yrs (A)& $10^4-10^7$& & $\sim 0.5$ & Bednarek (2003b) \\
                       &     & $Fe-\gamma\rightarrow n$    & $\sim 10^{6}$ yrs (A)& $\lesssim 10^6$& $\propto$ E$^{-3.1}$ & $\sim 4$ & Anchordoqui et al. (2003a)\\
\hline
{\bf binary systems} & & & && & & \\
A0535+26 & 2.6       & $pp$ &$\sim 10^{6}$ yrs (A)& $3\cdot 10^2-10^3$ & $4.2\cdot 10^{-3}$E$^{-2.35}$& few & Anchordoqui et al. (2003b) \\
\hline
{\bf Microquasars}   & 1-10& $pp$ & days/yrs (D) & $10^3-10^5$ &  $\propto$ E$^{-2}$& 1-300 & Distefano et al. (2002) \\
\hline
{\bf magnetars}      & 3-16& $p\gamma$ &$< 10^{4}$ yrs (A) 
& $\lesssim 100$ TeV & $\propto$ E$^{-2}$& $1.7 \times$ (0.1/$\Delta \Omega)(5/d)^2$ & Zhang et al. (2003) \\
\hline
\hline
\end{tabular} 
\label{tab1}
\end{sidewaystable}

\end{document}